\documentclass[a4paper,11pt]{article}
\pdfoutput=1

\usepackage{jheppub1}
\usepackage[T1]{fontenc}
\usepackage{amssymb}
\usepackage{graphicx}
\usepackage{amsmath}
\usepackage{hyperref}
\usepackage{tensor}
\usepackage{epstopdf}
\usepackage{extarrows}
\newcommand{\td}{\text{d}}
\newcommand{\el}{\ell_{\text{AdS}}}
\newcommand{\C}{\mathcal{C}}

\def \tf {\mathfrak{f}}
\def \R {\mathfrak{R}}
\def \pD {\mathfrak{D}}
\def \h {\mathfrak{h}}

\begin{document}
\title{Upper bound on cross-sections inside black holes and complexity growth rate}
\author{Run-Qiu Yang}
\emailAdd{aqiu@tju.edu.cn}

\affiliation{Center for Joint Quantum Studies and Department of Physics, School of Science, Tianjin University, Yaguan Road 135, Jinnan District, 300350 Tianjin, P.~R.~China}

%
\abstract{
This paper studies cross-sections inside black holes and conjectures a universal inequality: in a static $(d+1)$-dimensional asymptotically planar/spherical Schwarzschild-AdS spacetime of given energy $E$ and AdS radius $\el$, the ``size of cross-section'' inside black holes is bounded by $8\pi E\el/(d-1)$. To support this conjecture, it gives the proofs for cases with spherical/planar symmetries and some special cases without planar/spherical symmetries. As one corollary, it shows that the complexity growth rate in complexity-volume conjecture satisfies the upper bound argued by quantum information theory. This makes a first step towards proving the conjecture that the vacuum black hole has fastest complexity growth in the systems of same energy. It also finds a similar bound for asymptotically flat black holes, which gives us an estimation on the largest interior volume of a large evaporating black hole.
}
\maketitle
\noindent
\section{Introduction}\label{intro}
Black holes, as ultra dense objects in universe, exhibit many fascinating physical and mathematical properties. Particularly, many such properties can be presented by some universal inequalities, such as the positive mass theorem~\cite{Schoen1979,Schoen1981}, the second law of black holes~\cite{Wald_2001,Bardeen1973}, the Penrose inequality~\cite{Bray:2003ns,Huisken2001,Bray2001,Mars_2009,Lee2015} and so on.
Most of these universal inequalities focus on the horizon and its exterior. However, the recent developments suggested that the inner structures of black holes may also play important roles in considering the black hole physics. For example, the interior of black hole play crucial role in holographic computational complexity~\cite{Susskind:2014rva,Stanford:2014jda,Alishahiha:2015rta,Brown:2015bva,Brown:2015lvg,Bhattacharya:2019zkb,Carmi:2017jqz} and recent developments toward the resolution of information paradox in holography~\cite{almheiri2019page,Penington:2019npb,almheiri2019entropy,chen2019information}. The interior is important in the proposal of ``quantum Penrose inequality''~\cite{Bousso_2019,Bousso_2019b}. It has been suggested that the volume of interior of black hole may be relevant to the information paradox~\cite{Christodoulou:2014yia,Bengtsson:2015zda,Ong:2015tua,Mathur:2020ely}.
The universal inequalities about the inner structures of black holes are still lack of exploring.

This paper makes a first step to explore a new universal inequality about inner geometry of stationary black holes and exhibits a few of applications in black hole physics. The inequality arises from following simple question. Consider a $(d+1)$-dimensional Schwarzschild-AdS black hole, of which the metric reads
\begin{equation}\label{defSW1}
  \td s^2=\frac1{z^2}\left[-f\td t^2+\frac{\td z^2}{f}+\delta_{ij}\td x^i\td x^j\right]\,.
\end{equation}
Here $f=1/\el^2-f_0z^d$ and $\el$ is the AdS radius.
Inside horizon, $z$ is time but $t$ is spatial coordinate. For a class of special slices which are fixed ``time'' $z$, the volume reads $V=\int\td t\Sigma$ with $\Sigma=V_{d-1}\sqrt{-f}z^{-d}$.
Here $V_{d-1}:=\int\td^{d-1}x$. Geometrically, $\Sigma$ can be interpreted as the ``size'' of a cross-section since its integration with respect to $t$ gives us the volume of this slice. Different $z$ will give us different size of cross-section. The directly computation shows that
\begin{equation}\label{conjbd1}
  \Sigma\leq 8\pi E\el/(d-1)\,.
\end{equation}
Here $E$ is the total energy/mass of the spacetime.
This paper conjectures that, for a stationary black hole, if (i) outermost horizon is connected Killing horizon and has positive surface gravity, (ii) the spacetime is \textit{asymptotically} spherical/planar Schwarzschild-AdS~\footnote{This requirement is stronger than ``asymptotically AdS'', see Ref.~\cite{shi2018regularity} for a rigours mathematical definition about what is ``asymptotically Schwarzschild-AdS''.}\cite{shi2018regularity},  and (iii) dominate energy condition and Einstein equation are satisfied, then inequality~\eqref{conjbd1} is always true.
To support this conjecture, this paper gives the proofs on some situations which cover most of physical interesting cases.

Though it is not the original motivation of inequality~\eqref{conjbd1}, this paper finds that the inequality~\eqref{conjbd1} has important application in holographic duality. It has been argued from quantum information theory that the complexity growth rate $\dot{\C}(\tau)$ satisfies Lloyd's bound~\cite{Lloyd2000}
\begin{equation}\label{Lloydbd1}
  \dot{\C}(\tau)\leq2E/\pi\,.
\end{equation}
Here $\C(\tau)$ is the complexity of a time-dependent system and $E$ is the total energy. This bound describes the ultimate speed of quantum computations~\cite{Lloyd2000}.
We will show that, if the inequality~\eqref{conjbd1} is true, then the complexity growth rate of stationary black hole in ``complexity-volume'' (CV) conjecture~\cite{Susskind:2014rva,Stanford:2014jda} always satisfies the Lloyd's bound~\eqref{Lloydbd1}.
Any regular matter (satisfies dominate energy condition) in the bulk always slows down the complexity growth. This matches with a conjecture that black hole has fastest information scramnbling~\cite{Sekino:2008he}. The similar bound of growth rate was once conjectured in ``complexity=action'' (CA) conjecture~\cite{Brown:2015bva,Brown:2015lvg} but has been found to be violated even in Schwarzschild black holes~\cite{Carmi:2017jqz,Kim:2017qrq,HosseiniMansoori:2018gdu,Mahapatra:2018gig}. The inequality~\eqref{conjbd1} thus gives us a new viewpoint to compare CV and CA conjectures.

It will also show that the inequality~\eqref{conjbd1} has a generalization asymptotic flat spacetimes $\Sigma\leq c_dE^{(d-1)/(d-2)}$ with a dimension-dependent positive number $c_d$. We find that this bound is relevant to ``interior volume'' of black hole proposed by Ref.~\cite{Christodoulou:2014yia} and can give us an estimation on the possible largest exterior volume of an evaporating black hole.

\section{Cross-section inside the black hole}\label{csibh}
We first clarify the precise definition of ``size of cross-section''. Consider a $(d+1)$-dimensional stationary black hole with an outermost non-degenerated connected Killing horizon. Assume $\xi^I$ to be Killing vector field which is timelike outside.  A cross-section $\mathcal{S}_{d-1}$ is an arbitrary spacelike $(d-1)$-dimensional submanifold inside the black hole (If there are inner horizons, then ``inside black hole'' means the region between the outermost horizon and next-outermost horizon). The size of this cross-section is defined as
\begin{equation}\label{definSigma1}
  \Sigma[\xi^I, \mathcal{S}_{d-1}]:=\int_{\mathcal{S}_{d-1}}\xi^In^J\td\Sigma_{IJ}\,.
\end{equation}
Here $\td\Sigma_{IJ}$ is the outer-past directed surface element of $\mathcal{S}_{d-1}$ and $n_I$ is a unit normal covector of $\mathcal{S}_{d-1}$ which satisfies $n^I\xi_I=0$.
Geometrically, $\Sigma$ stands for the flux of vector field $\xi^I$ in the surface $\mathcal{S}_{d-1}$.
The cross-section $\mathcal{S}_{d-1}$ is trivial if $\xi^I$ tangent to $\mathcal{S}_{d-1}$ as the size is zero. For nontrivial cross-section, $n_I$ is the unique future-directed time-like normal covector.
We only consider the maximally extended cross-section, i.e., the cross-section which is not a real subset of any other cross-section.

In a general stationary $(d+1)$-dimensional spacetime, the metric can locally be expressed as
\begin{equation}\label{metricflat1}
\begin{split}
  \td s^2=&\frac1{z^2}[-f\td t^2+\chi\td z^2+2v_i\td t\td x^i+h_{ij}\td x^i\td x^j]\,
  \end{split}
\end{equation}
with $(\partial/\partial t)^I=\xi^I$. The functions $f,\chi,v_i$ and $h_{ij}$ may depend on $\{z,x^i\}$ but do not depend on $t$. A nontrivial cross-section  $\mathcal{S}_{d-1}$ can be parameterized by $z=z_{\mathcal{S}}(x^i)$ and $t=t_{\mathcal{S}}(x^i)$. The position of cross-section depends on the choice of $t_S(x^i)$, however, it is shown in appendix~\ref{matrix1} that its size $\Sigma[\xi^I,\mathcal{S}_{d-1}]$ is independent of the choice on $t_{\mathcal{S}}(x^i)$. Based on this property, we can compute $\Sigma[\xi^I,\mathcal{S}_{d-1}]$ by choosing $t_{\mathcal{S}}(x^i)=0$ and find
\begin{equation}\label{defLRL}
  \Sigma[\xi^I,\mathcal{S}_{d-1}]=\int_{z=z_{\mathcal{S}}(x^i)} z^{-d}\sqrt{-f-|v|^2}\sqrt{\tilde{h}}\td^{d-1}x\,.
\end{equation}
Here $\tilde{h}:=$det$(\tilde{h}_{ij}), |v|^2=\tilde{h}^{ij}v_iv_i$ and
$$\tilde{h}_{ij}:=h_{ij}+\chi\partial_iz_S\partial_jz_S$$
is the induced metric of cross-section $\{t=0,z=z_{\mathcal{S}}(x^i)\}$. See the appendix~\ref{matrix1} for mathematical proof on Eq~\eqref{defLRL}.


Alternatively, we can choose the Bondi-Scahs coordinates and metric reads~\cite{Sachs1962,M_dler_2016}
\begin{equation}\label{metricBS1}
\begin{split}
  \td s^2&=\frac1{z^2}[-\mathfrak{f}e^{2\beta}\td u^2+2e^{2\beta}\td u\td z\\
  &+q_{ij}(\td x^i-U^i\td u)(\td x^j-U^j\td u)]
  \end{split}
\end{equation}
with $(\partial/\partial u)^I=\xi^I$ and gauge $\partial_zq=0$. The functions $\tf, \beta, U^i$ and $q_{ij}$ may depend on $\{z,x^i\}$ but do not depend on $u$. A general cross-section is parameterized by $z=z_{\mathcal{S}}(x^i)$ and $u=u_{\mathcal{S}}(x^i)$. It is still true that $\Sigma[\xi^I,\mathcal{S}_{d-1}]$ is independent of the choice on $u_{\mathcal{S}}(x^i)$ and so we can compute $\Sigma[\xi^I,\mathcal{S}_{d-1}]$ by choosing $u_S(x^i)=0$. Then we find
\begin{equation}\label{defLRL2}
  \begin{split}
  &\Sigma[\mathcal{S}_{d-1},\xi^I]\\
  =&\int_{z=z_{\mathcal{S}}(x^i)}z^{-d}e^{\beta}\sqrt{-\tf-e^{2\beta}|\partial z|^2+2U^i\partial_iz}\td V_{d-1}\,.
  \end{split}\,.
\end{equation}
Here $\td V_{d-1}:=\sqrt{q}\td^{d-1}x$ and $|\partial z|^2=q^{ij}\partial_iz\partial_jz$. See the appendix~\ref{bfg1} for mathematical proof on Eq~\eqref{defLRL2}. For arbitrary positive function $p(x^i)$, the coordinates transformation $\{z\rightarrow zp(x^i)\}$ changes one allowed Bondi-Scahs coordinates into an other. By this freedom, we can set the cross-section to have constant $z$. Under this gauge choice, we have a simple formula,
\begin{equation}\label{defLRL2b}
  \Sigma=\int_{z=\bar{z}} z^{-d}e^{\beta}\sqrt{-\tf}\td V_{d-1}\,.
\end{equation}
This will useful when we prove our inequality.

\section{Examples in some special cases}
Before we discuss and try to prove our bound in general cases, it would be worthy of showing some examples. These examples may give readers an intuition of generality about this bound.

In the first example we consider Schwarzschild-AdS black hole. We here give a detailed computation for the inequality~\eqref{conjbd1}. The metric of a planar AdS-Schwarzschild black hole is given by Eq.~\eqref{defSW1} with $f=1/\el^2-f_0z^d$. Here $f_0$ is a positive parameter. The horizon then locates at $z_h=(f_0\el^2)^{-1/d}$. The mass of the black hole is
\begin{equation}\label{defMass}
  M=\frac{(d-1)V_{d-1}}{16\pi}f_0\,.
\end{equation}
Due to planar symmetry, the maximal cross-section must have a constant $z$ and so we have
\begin{equation}\label{maxsigmas1}
\begin{split}
  &\max\Sigma=\max V_{d-1}\sqrt{P_d(z)}=V_{d-1}\sqrt{\max P_d(z)},\\
  &P_d(z):=-f(z)z^{-2d}\,.
  \end{split}
\end{equation}
The maximal value of $P_d(z)$ is determined by following equation
\begin{equation}\label{saddlef1}
  P_d'(z)=0\Rightarrow f'(z)z-2df(z)=0\,.
\end{equation}
For Schwarzschild black hole $f(z)=1/\el^2-f_0z^d$, one can find that the solution of Eq.~\eqref{saddlef1} reads $z^d=z_m^d:=2/(f_0\el^2)$ and so we find
\begin{equation}\label{maxFs1}
  \max P_d(z)=P_d(z_m)=\frac{\el^2f_0^2}4\,.
\end{equation}
Then we obtain the inequality~\eqref{conjbd1} for a planar AdS-Schwarzschild black hole. We see that $\Sigma$ can saturate the upper bound in a planar Schwarzschild black hole.

The case will be a little complicated if we consider the spherically or hyperbolically symmetric AdS-Schwarzschild black hole. The function $f$ will be
$$f=kz^2+\frac1{\el^2}-f_0z^d\,,$$
The Eq.~\eqref{maxsigmas1} becomes
\begin{equation}\label{maxsigmas2}
\begin{split}
  &\max\Sigma=V_{d-1}\sqrt{\max P_d(z)}\\
  &P_d(z):=-[kz^2+\frac1{\el^2}-f_0z^d]z^{-2d}\,.
  \end{split}
\end{equation}
The solution of $P_d'(z)=0$ now satisfies
\begin{equation}\label{solufz1}
  z_m^d=\frac{2}{\el^2f_0}+\frac{2(d-1)k}{df_0}z_m^2>0\,.
\end{equation}
We take this expression into~\eqref{maxsigmas2} and eliminate $z^d$ and $z^{-2d}$ terms. After some algebras, we find
\begin{equation}\label{maxpzk}
\begin{split}
  &\max P_d(z)-\frac{\el^2f_0^2}4=P_d(z_m)-\frac{\el^2f_0^2}4\\
  &=-\frac{\el^2f_0^2}4\frac{[kz_m^2\el^2(d-1)^2+d^2]kz^2_m\el^2}{(dk\el^2z_m^2+d-k\el^2z_m^2)^2}\,.
  \end{split}
\end{equation}
The dominate energy condition implies $f_0\geq0$, so Eq.~\eqref{solufz1} implies
$$kz_m^2\el^2(d-1)^2>-(d-1)>-d^2\,.$$
If $k>0$, we see that $\max P_d(z)-\el^2f_0^2/4<0$. Thus, in spherically symmetric AdS-Schwarzschild black hole, the inequality~\eqref{conjbd1} is still true but upper bound cannot be attained. If $k=-1$, we see that $\max P_d(z)-\el^2f_0^2/4>0$. Thus, the hyperbolically symmetric black hole violates our bound. This is the reason why we require the spacetime should have asymptotically spherical/planar symmetry.

In the second example, we consider planar or spherical Reissner-Nordstr\"{o}m AdS (RN-AdS) black hole. The metric is still given by Eq.~\eqref{defSW1} but now the function $f$ is given by
\begin{equation}\label{RNfz1}
  f=kz^2+1/\el^2-f_0z^d+\tilde{q} z^{2d-2}\,.
\end{equation}
Here $\tilde{q}:=q^2\geq0$ is the charge parameter. The mass of this black is still given by Eq.~\eqref{defMass}. The maximal cross-section is still given by Eq.~\eqref{maxsigmas1} but $f$ now is replaced by Eq.~\eqref{RNfz1} and so $P_d=P_d(z,\tilde{q})$. Assume that $z_m(\tilde{q})$ is the point which maximizes $P_d$. We then have
\begin{equation}\label{onshellsigma}
  f'(z_m,\tilde{q})z_m-2df(z_m,\tilde{q})=0
\end{equation}
and
\begin{equation}\label{onshellPm}
  \max P_d=\mathcal{P}(\tilde{q}):=-f(z_m(\tilde{q}),\tilde{q})z_m^{-2d}(\tilde{q})\,.
\end{equation}
Here $f'(z,\tilde{q})=\partial_zf(z,\tilde{q})$. It would be a little complicated to solve Eq.~\eqref{onshellsigma} and compute Eq.~\eqref{onshellPm} directly. We compute the value of $\td \mathcal{P}(\tilde{q})/\td\tilde{q}$
\begin{equation}\label{dpdzeq1}
  \begin{split}
  &\td \mathcal{P}(\tilde{q})/\td\tilde{q}=-\partial_{\tilde{q}}f|_{z_m=z_m(\tilde{q})}z_m(\tilde{q})^{-2d}\\
  &-[f'(z_m(\tilde{q}),\tilde{q})z_m(\tilde{q})-2df(z_m(\tilde{q}),\tilde{q})] z_m(\tilde{q})^{-2d-1}\frac{\td z_m(\tilde{q})}{\td\tilde{q}}\,.
  \end{split}
\end{equation}
Noting the fact that $z_m(\tilde{q})$ satisfies Eq.~\eqref{onshellsigma}, we obtain a simple result
\begin{equation}\label{dPdq1}
  \td \mathcal{P}(\tilde{q})/\td\tilde{q}=-z_m(\tilde{q})^{-2d}\partial_{\tilde{q}}f|_{z_m=z_m(\tilde{q})}=-z_m^{-2}(\tilde{q})<0\,.
\end{equation}
When $\tilde{q}=0$, the black hole is just planar or spherical Schwarzschild black hole and we have $\mathcal{P}(0)\leq\frac{\el^2f_0^2}{4}$. Thus, we find that
$$\forall~\tilde{q}\geq0,~~\mathcal{P}(\tilde{q})\leq \mathcal{P}(0)\leq\frac{\el^2f_0^2}{4}\,.$$
This shows that the inequality~\eqref{conjbd1} is still true for RN-AdS black hole and the saturation can appear only when $\tilde{q}=k=0$.

We note that, though we should restrict $\tilde{q}$ to be nonnegative in physics, it is still a solution of Einstein's equation when $\tilde{q}<0$, i.e. we replace $q\rightarrow iq$. Such black hole is a solution for the Einstein-Maxwell system with a phantom coupling of Maxwell field, i.e. Maxwell field minimally couples to gravity with ``wrong'' sign. If we take a negative value for $\tilde{q}$, then we see that $\mathcal{P}(\tilde{q})>\mathcal{P}(0)$ and our inequality~\eqref{conjbd1} may be violated. We note that such black hole violates dominate energy condition. This example implies that we should propose a suitable energy condition as a necessary condition for the inequality~\eqref{conjbd1}.

In the third example, we consider a non-static stationary black hole. The simplest one is the BTZ black hole, of which the metric reads~\cite{Banados:1992wn,Banados:1992gq}
\begin{equation}\label{metricBTZ}
  \td s^2=\frac1{z^2}\left[-\tilde{f}(z)\td t^2+\frac{\td z^2}{\tilde{f}(z)}+(\td\phi-Jz^2\td t/2)^2\right]\,
\end{equation}
with $\tilde{f}(z)=1/\el^2-f_0z^2+J^2z^4/4$. Comparing with Eq.~\eqref{metricflat1} we see
\begin{equation}\label{ffchieq1}
  v_i=v=Jz^2/2,~~f=\tilde{f}-J^2z^4/4=\tilde{f}-v^2\,.
\end{equation}
A general cross-section between the outmost horizon and inner horizon is a line and can be defined by $\{z=z_S(\phi),t=t_S(\phi)\}$. Using Eq.~\eqref{defLRL}, the size of a cross-section is given by
\begin{equation}\label{btzsig1}
\begin{split}
  \Sigma&=\int_{z=z_S(\phi)}z^{-2}\sqrt{-f-v^2}\td\phi\\
  &=\int z_S^{-2}\sqrt{f_0z_S^2-1/\el^2-J^2z_S^4/4}\td\phi\\
  &=\int \sqrt{P_2(z_S)-J^2/4}\td\phi
  \end{split}
\end{equation}
Here $P_d(z)=(f_0z^d-1/\el^2)z^{-2d}$. The mass of this BTZ black hole reads $M=V_1 f_0/16\pi$ with $V_2=2\pi$. It is easy to find
\begin{equation}\label{btzsig2}
\begin{split}
  \max\Sigma&=\int \sqrt{\el^2f_0^2/4-J^2/4}\td\phi\\
  &=V_1\sqrt{\el^2f_0^2/4-J^2/4}\leq8\pi\el M\,.
  \end{split}
\end{equation}
We see the the BTZ black hole also satisfies our inequality~\eqref{conjbd1} and the bound is saturated only when $J=0$.

\section{Relationship to complexity and maximal interior volume of black hole}
Though it is completely based on geometrical considerations, the ``size of cross-section'' has directly relationship to the complexity growth rate in CV conjecture.
The CV conjecture (see Refs.~\cite{Susskind:2014rva,Stanford:2014jda,Alishahiha:2015rta} for more details) states that the complexity of a boundary state in asymptotically AdS spacetime is proportional to the maximal volume of space-like codimension-one surface $W_d$ connecting boundary time slices $S_L(t_L)$ and $S_R(t_R)$, i.e.,
\begin{equation}\label{CV}
  \mathcal{C}=\max_{\partial W_d=S_L(t_L)\cup S_R(t_R)}\frac{V[W_d]}{G_N \ell}\,,
\end{equation}
Here $G_N$ is the Newton constant and we set $G_N=1$ for convenience, $\ell$ is a length scale associated to the bulk geometry such as the horizon radius or AdS radius and so on. We take $\ell=4\pi^2\el/(d-1)$.\footnote{We choose $\ell$ in this way to match the Lloyd bound~\eqref{Lloydbd1}. Note that the complexity always has the freedom in defining ``unit complexity'' and so Lloyd bound~\eqref{Lloydbd1} itself has freedom upon a universal factor. }

\begin{figure}
  \centering
  \includegraphics[width=.3\textwidth]{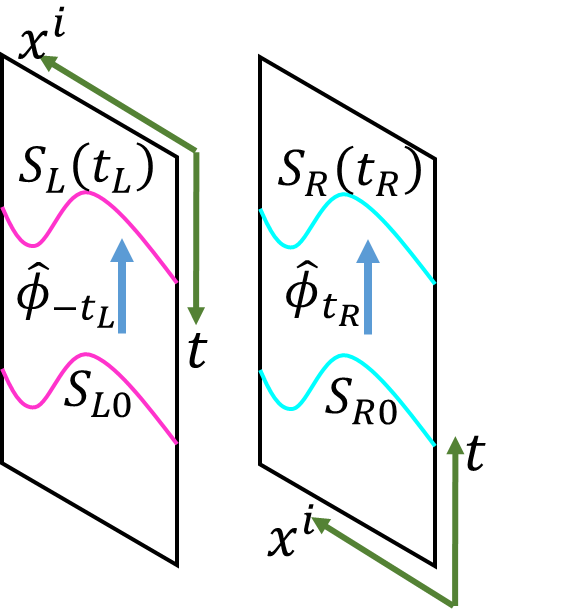}
  \includegraphics[width=.43\textwidth]{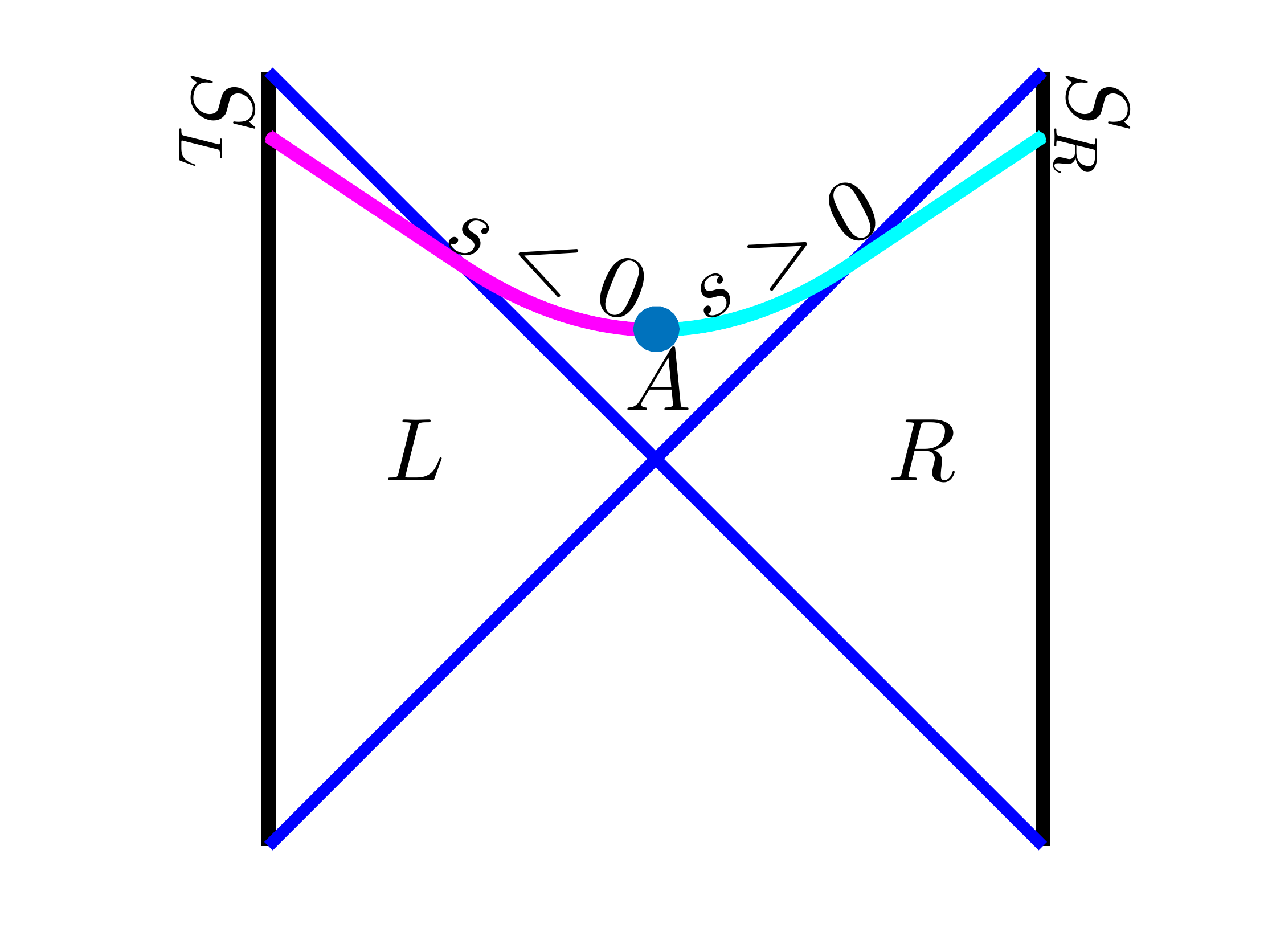}
  \caption{The schematic diagram about the boundary slices  (left) and extremal surface $W_d$ (right). In the right panel, the possible inner horizons and singularities are irrelevant, so they are not showed. For every fixed $x^i$, the time (i.e. coordinate time $t$) of two boundary time slices will always have opposite sign.}
   \label{CVinSAdS1}
\end{figure}

We choose coordinates gauge~\eqref{metricflat1}. The right boundary slice is given by $S_R(t_R):=\hat{\phi}_{t_R}S_{R0}$, where $S_{R0}$ is initial boundary slice defined by $z=0$ and $t=T(x^i)$ with arbitrary function $T(x^i)$. Here $\hat{\phi}_\lambda$ is 1-parameter group of diffeomorphisms generated by timelike Killing vector $(\partial/\partial t)^I$ at the boundary. The left boundary slice is given by $S_L(t_L):=\hat{\phi}_{-t_L}S_{L0}$ and $S_{L0}$ is defined by $t=-T(x^i)$. In this choice, for every fixed $x^i$, the coordinate time of two boundary slices always have opposite sign. We can parameterize $W_d$ by $z=z(s,x^i)$ and $t=t(s,x^i)$. See Fig.~\ref{CVinSAdS1}.
We use the Penrose diagram of planar/spherical symmetric black hole as example in Fig.~\ref{CVinSAdS1} for visualization, however, the computations and arguments can be applied into arbitrary stationary AdS black holes. Two boundary slices $S_L$ and $S_R$ are given by $s=\pm\infty$, i.e.
\begin{equation}\label{twoslinces}
\begin{split}
  S_L&=\{z=0,t=t(-\infty,x^i)\}\,,\\
  S_R&=\{z=0,t=t(\infty,x^i)\}\,.
  \end{split}
\end{equation}
and we have relationship $t(-\infty,x^i)=-t(\infty,x^i)$.

As the spacetime is stationary, the maximal volume depends on only the value of $t_L+t_R$
and we only need to consider the symmetric time slices, i.e., $t_L=t_R$. In this case, the extremal surface contains two parts ($s<0$ and $s>0$). Noting the boundary condition~\eqref{twoslinces} and the fact that bulk geometry of one side is just the copy of the other, we find that there is following relationship
\begin{equation}\label{symmwd1}
  t(s,x^i)=-t(-s,x^i), z(s,x^i)=z(-s,x^i)
\end{equation}
if $W_d$ is an extremal hypersurface. The intersections of two parts (i.e. the parts of $s\geq0$ and $s\leq0$) is denoted by $A$, which is given by $t(0,x^i)=0$ and $z=z(0,x^i)=z_A(x^i)$.

The induced metric on $W_d$ reads
\begin{equation}\label{inducedWd1}
\begin{split}
  \td s^2_W=&\frac1{z^2}[-f(t'\td s+\partial_it\td x^i)^2+\chi(z'\td s+\partial_iz\td x^i)^2\\
  &+2v_i(t'\td s+\partial_jt\td x^j)\td x^i+h_{ij}\td x^i\td x^j]\,.
  \end{split}
\end{equation}
Here ``$~'~$'' stand for the partial derivative with respect to $s$. Now we define
\begin{equation}\label{defNN1}
  N=-ft'^2+\chi z'^2,~N_i=v_it'+\chi z'\partial_iz-ft'\partial_it\,,
\end{equation}
and
\begin{equation}\label{defh1}
  \h_{ij}=-f\partial_it\partial_jt+\chi\partial_iz\partial_jz+2v_{(i}\partial_{j)}t+h_{ij}
\end{equation}
Then we have
$$\td s_W^2=\frac1{z^2}[N\td s^2+2N_i\td s\td x^j+\h_{ij}\td x^i\td x^j]\,.$$
The volume functional of $W_d$ now reads
\begin{equation}\label{volumeW1}
  V[W_d]=\int z^{-d}\sqrt{N-\h^{ij}N_iN_j}\sqrt{\h}\td^{d-1}x\td s\,.
\end{equation}
Note that only $N$ and $N_i$ depend on the value of $t'$. The above volume functional is an analog of action functional. The two variables $t(s,x^i)$ and $z(s,x^i)$ are two ``fields'' and parameter $s$ plays the role of ``time''. Thus the canonical momentum conjugate to ``field''  $t$ reads
\begin{equation}\label{mometumt1}
\begin{split}
  \mathcal{P}(s_0)&=\int_{s=s_0} \frac{\partial}{\partial t'}\left[\sqrt{N-\h^{ij}N_iN_j}\frac{\sqrt{\h}}{z^d}\right]\td^{d-1}x\\
  =&\int_{s=s_0}\frac{\frac{\partial N}{\partial t'}-2\h^{ij}N_i\frac{\partial N_j}{\partial t'}}{2\sqrt{N-\h^{ij}N_iN_j}}\frac{\sqrt{\h}}{z^d}\td^{d-1}x\,.
  \end{split}
\end{equation}
From Eqs.~\eqref{defNN1} we can find that
\begin{equation}\label{valuedNdt1}
  \frac{\partial N}{\partial t'}=-2ft',~~\frac{\partial N_j}{\partial t'}=-f\partial_it+v_j\,.
\end{equation}
Thus, we have
\begin{equation}\label{mometumt2}
  \mathcal{P}(s_0)=\int_{s=s_0}\frac{-ft'-\h^{ij}N_i(v_j-f\partial_it)}{\sqrt{N-\h^{ij}N_iN_j}}\frac{\sqrt{\h}}{z^d}\td^{d-1}x
\end{equation}
The extremal surface is obtained by Euler-Lagrangian equation of volume functional~\eqref{volumeW1}. The maximal volume $V$, i.e., the on-shell value of $V$, is only the function $t_L+t_R=\tau$ and so we have $V_{\text{on-shell}}=V_{\text{on-shell}}(\tau)$. The growth rate then reads
\begin{equation}\label{dvdteq1}
  \ell\dot{\C}=\frac{\td V_{\text{on-shell}}}{\td\tau}=\left.\frac{\partial V_{\text{on-shell}}}{\partial t_R}\right|_{\text{fix}~t_L}\,.
\end{equation}
As the partial derivative of on-shell action with respect to canonical variable $t$ gives us the canonical momentum, we see
\begin{equation}\label{cdotp1}
  \ell\dot{\C}= \mathcal{P}(\infty),
\end{equation}
As the volume functional does not depend on $t$ explicitly, $\mathcal{P}(s)$ will be independent of $s$ and so
\begin{equation}\label{ps0p0}
  \mathcal{P}(\infty)=\mathcal{P}(s)\,.
\end{equation}
We can compute $\mathcal{P}(\infty)$ at $s=0$, i.e at the surface $A$. Eq.~\eqref{symmwd1} implies $z'|_{A}=0$ and $\partial_it|_A=0$, so we find
\begin{equation}\label{valueNNi1}
  \h_{ij}|_A=\tilde{h}_{ij},~~N|_A=-ft'^2,~~N_i|_A=v_it'\,.
\end{equation}
Thus
\begin{equation}\label{odefort1new1b}
  \ell\dot{\C}=\mathcal{P}(0)=\int_{A} z^{-d}\sqrt{-f-|v|^2}\sqrt{\tilde{h}}\td^{d-1}x=\Sigma[\xi^I,A]\,.
\end{equation}
In RN-AdS black holes, it recovers the result reported by Ref.~\cite{Carmi:2017jqz} after we specify $S_{L0}$ and $S_{R0}$ to be equal-$t$ slice of boundary.
We see that the complexity growth rate is given by size of cross-section $A$ and so Eq.~\eqref{Lloydbd1} suggests Eq.~\eqref{conjbd1}. This offers a way to argue Eq.~\eqref{conjbd1} by AdS/CFT correspondence and information theory. In turn, the proof of Eq.~\eqref{conjbd1} is significant for CV conjecture.

If we move the boundary slices $S_L$ and $S_R$ into the horizon, then we find that $\hat{\phi}_\lambda$ is the tangent map of horizon. The $V_{\text{on-shell}}(\tau)$ becomes the ``the maximal interior volume'' attached by the horizon slices $S_L$ and $S_R$, which captures the idea of ``how much space is inside''~\cite{Christodoulou:2014yia}. This quantities is suggested to be relevant to the information paradox~\cite{Christodoulou:2014yia,Bengtsson:2015zda,Ong:2015tua}. From the above discussion, it is clear that the growth rate of such volume is still given by size of cross-section $A$.

\section{Proofs on the inequality}
In this section, we will give proof of our bound. We use the Bondi-Scahs coordinates~\eqref{metricBS1}. As we have argued at the end of Sec.~\ref{csibh}, we can always choose the Bondi-Scahs coordinates suitably so that the maximal cross-section is $z=\bar{z}$ with a constant $\bar{z}$. Using Cauchy-Schwartz inequality and defining $F(z):=V_{d-1}^{-1}\int(-z^{-2d}\tf)\td V_{d-1}$, we have
\begin{equation}\label{pertub2}
 \max\Sigma^2\leq V_{d-1}F(\bar{z})\int_{z=\bar{z}}e^{2\beta}\td V_{d-1}\,.
\end{equation}
%
Here $V_{d-1}:=\int \td V_{d-1}$. Note that $\td V_{d-1}$ and $V_{d-1}$ are independent of $z$ due to the gauge $\partial_zq=0$. The requirement (ii) implies the boundary conditions $\beta|_{z=0}=0$.



Assume that $T_{IJ}$ is the energy momentum tensor of matters. The Einstein's equation shows that~\cite{M_dler_2016}
\begin{equation}\label{uucompteq}
  \partial_z\beta=-\frac{zq^{ik}q^{jl}(\partial_zq_{kl})\partial_zq_{ij}}{8(d-1)}-\frac{4\pi z}{d-1} {T}_{zz}\,.
\end{equation}
and
\begin{equation}\label{urcompteq4}
\begin{split}
  &-(d-1)z^{d-1}\partial_z(z^{-d}{\tf})\\
  &=e^{2\beta}[\R+2(\pD\beta)^2]-\pD^2e^{2\beta}-\frac{e^{-2\beta}}{2}q_{ij}\partial_zU^i\partial_zU^j\\
  &+\frac{d(d-1)e^{2\beta}}{z^2\el^2}-z^{2d-2}\pD_i[\partial_z(U^i/z^{2d-2})]\\
  &-8\pi z^{-2}e^{2\beta}(\rho-P)\,.
  \end{split}
\end{equation}
Here $\R$ and $\pD_i$ are the scalar curvature and covariant derivative of $q_{ij}$. $T_{zz}$ is the $zz$ (null-null) component of $T_{IJ}$, $\rho=T_{IJ}n^In^J$ and $P=T_{IJ}m^Im^J$, where $n^I$ and $m^I$ are orthogonal future-directed timelike and outward spacelike vectors of subspace spanned by $\{x^i\}$. The dominate energy condition implies $T_{zz}\geq0$ and $\rho-P\geq0$.

We note that the first term in r.h.s. of Eq.~\eqref{uucompteq} is invariant under the coordinates transformation $x^i\rightarrow\tilde{x}^i=X^i(x)$, where $X^i(x)$ is independent of $z$. As $q^{ij}$ is the positive-defined metric of space spanned by $\{x^i\}$, we can find suitable coordinates transformation function $X^i_p$ at every point $p$ so that $q^{ij}|_p$ becomes diag$\{\lambda_1(p),\lambda_2(p),\cdots,\lambda_{d-1}(p)\}$ with $\lambda_i(p)>0$ (note that $\partial_zq_{kl}$ in this new coordinates may not be diagonalized) and so
\begin{equation}\label{firstqqz1}
  q^{ik}q^{jl}(\partial_zq_{kl})\partial_zq_{ij}|_p=\sum_{i,j}\lambda_i(p)\lambda_j(p)(\partial_zq_{ij})^2|_p\geq0\,.
\end{equation}
Then Eq.~\eqref{uucompteq} implies $\partial_z\beta\leq0$. Combining it with the boundary condition $\beta|_{z=0}=0$, we find $\beta\leq0$ and so
\begin{equation}\label{ineqFs1}
  \max\Sigma^2\leq V_{d-1}^2F(\bar{z})\,.
\end{equation}
After we integrate Eq.~\eqref{urcompteq4} on the transverse directions and neglect the boundary terms, we have
\begin{equation}\label{urcompteq7}
\begin{split}
  &\frac{\td}{\td z}(z^dF)=\frac{de^{2\beta}}{z^{d+1}\el^2}\\
  &+\frac{\int e^{2\beta}[\R+2({\pD}\beta)^2-q_{ij}A^iA^j-Q^2]\td V_{d-1}}{(d-1)z^{d-1}V_{d-1}}\,.
  \end{split}
\end{equation}
Here $A^i=e^{-\beta}\partial_zU^i/\sqrt{2}$ and $Q=\sqrt{8\pi(\rho-P)}e^{\beta}/z$. We then focus on following three cases.

The first case assumes that the spacetime is planar/spherically symmetric or deviates from such background only a little. Then the quantities $\partial_i\beta$ has order $\mathcal{O}(\epsilon)$ and the transverse metric becomes $q_{ij}=q_{ij}^{(0)}+\epsilon q_{ij}^{(1)}(z,x^i)$. The gauge $\partial_zq=0$ implies $q^{(0)ij}q^{(1)}_{ij}=\mathcal{O}(\epsilon^2)$. We then find that
\begin{equation}\label{pertureq2}
  e^{2\beta}[\R+2({\pD}\beta)^2]=(d-1)(d-2)e^{2\beta}k+\epsilon\mathring{\pD}^iY_i+\mathcal{O}(\epsilon^2).
\end{equation}
Here $\mathring{\pD}_i$ is the covariant derivative operator of $q_{ij}^{(0)}$ and $Y_i:=e^{2\beta}\mathring{\pD}^jq^{(1)}_{ij}$. Taking \eqref{pertureq2} into Eq.~\eqref{urcompteq7}, neglecting the boundary term, at the linear order of $\epsilon$, we have
\begin{equation}\label{urcompteq7b}
\begin{split}
  &\frac{\td}{\td z}(z^dF)=\frac{(d-2)ke^{2\beta}}{z^{d-1}}+\frac{de^{2\beta}}{z^{d+1}\el^2}\\
  &-\frac{\int e^{2\beta}[q_{ij}A^iA^j+Q^2]\td V_{d-1}}{(d-1)z^{d-1}V_{d-1}}+\mathcal{O}(\epsilon^2)\,.
  \end{split}
\end{equation}
Near the boundary $z\rightarrow0$, $\beta, A^i$ and $Q$ are required to decay fast enough, so when $z\rightarrow0$ we find
\begin{equation}\label{solutildeR0}
  F(z)\rightarrow-z^{-2d}[kz^2+1/\el^2-f_0z^d]=P_d(z)\,,
\end{equation}
where $f_0$ gives us mass according to Eq.~\eqref{defMass} and $P_d(z)$ is defined by Eq.~\eqref{maxsigmas2}. For finite $z$, Eq.~\eqref{urcompteq7b} implies
\begin{equation}\label{urcompteq7c1}
  \frac{\td}{\td z}(z^dF)\leq\frac{(d-2)k}{z^{d-1}}+\frac{d}{z^{d+1}\el^2}+\mathcal{O}(\epsilon^2)\,.
\end{equation}
Here we have used the fact $\beta\leq0$ and $k\geq0$. Integrating it and noting the asymptotically behavior \eqref{solutildeR0}, we find
\begin{equation}\label{solutildeR}
  F(z)\leq P_d(z)\,,
\end{equation}
We have shown that, in the case $k\geq0$, the maximal value of $P_d(z)$ is not larger than $f_0^2\el^2/4$, so Eq.~\eqref{ineqFs1} implies
\begin{equation}\label{upperforR}
  \max\Sigma\leq V_{d-1}f_0\el/2=\frac{8\pi E\el}{d-1}\,.
\end{equation}
Under the requirement $k\geq0$, above bound can be saturated only if $k=0$.

In the second case we consider the asymptotically planar Schwarzschild-AdS spacetimes when $d\geq4$. AdS/CFT duality conjectures that boundary is due to field theory in flat space. In many physical interesting cases, the energy momentum tensor of dual boundary field theory is assumed to decay rapidly enough when $x^i\rightarrow\infty$. In this case, we can find $q_{ij}\rightarrow\delta_{ij}, \partial_i\beta\rightarrow0$ when $x^i\rightarrow\pm\infty$. The integration in Eq.~\eqref{urcompteq7} is finite but $V_{d-1}\rightarrow\infty$. Thus the inequality~\eqref{urcompteq7b} is true and we still obtain inequality~\eqref{upperforR}. It needs to note that there are some physically interesting cases where the energy momentum tensor of dual boundary field are not decay when $x^i\rightarrow\infty$. It is not clear currently if our inequality will be still true.

In the third case, we assume that the spacetime is 3+1 dimensional.  On the $u=$constant null sheet, we denote $\Gamma_z$ to be a fixed $z$ 2-d spacelike surface, of which the metric is $q_{ij}(z,x)$. As $q_{ij}(z,x)$ is smooth between the maximal cross-section and asymptotic boundary, we find the surfaces $\{\Gamma_z|z\in[\bar{z},0)\}$ have a same topology. Because we require that the spacetime geometry is asymptotically planar/symmetric Schwarzschild-AdS black hole, the surface $\Gamma_z|_{z\rightarrow0}$ must be homeomorphic to a plane or a sphere. In 2-dimensional case, this means that all these surfaces in $\{\Gamma_z|z\in[\bar{z},0)\}$ are globally conformally to a plane or sphere.  As the result,  we can always find coordinates transformation $\{x^i\}\rightarrow\{y^i\}$ and a scalar $\Phi(z,y^i)$ suitably for every $\Gamma_z$ so that
\begin{equation}\label{conformetric}
  {q}_{ij}\td x^i\td x^j=e^{2\Phi}\gamma_{ij}^{(\tilde{k})}(y)\td y^i\td y^j\,
\end{equation}
with $\tilde{k}=1$ and 0. Here $\gamma_{ij}^{(0)}=\delta_{ij}$ and $\gamma_{ij}^{(1)}$ is the metric of a unite sphere.
Under the conformal transformation, we have
\begin{equation}\label{conformeq1}
\begin{split}
  &\int e^{2\beta}[\R+2({\pD}\beta)^2]\td V_{2}\\
  =&\int e^{2\beta}[2\tilde{k}-2\hat{\pD}^2{\Phi}+2(\hat{\pD}\beta)^2]\sqrt{\gamma^{(\tilde{k})}(y)}\td^2y\,.
  \end{split}
\end{equation}
Here $\hat{\pD}_i$ is the covariant derivative operator of conformal metric $\gamma_{ij}^{(\tilde{k})}$. Take it into Eq.~\eqref{urcompteq7} and we have
\begin{equation}\label{aurcompteq8b}
\begin{split}
  &\frac{\td}{\td z}(z^3F)=\frac{3e^{2\beta}}{z^4\el^2}+\frac{1}{z^2V_2}\int e^{2\beta}[\tilde{k}+(\hat{\pD}\beta)^2-\hat{\pD}^2{\Phi}\\
  &-\gamma^{(\tilde{k})}_{ij}\tilde{A}^i\tilde{A}^j-Q^2]\sqrt{\gamma^{(\tilde{k})}}\td^2y\,.
  \end{split}
\end{equation}
Here $\tilde{A}^i=e^{-\Phi}A^i$. $V_2$ is transversa volume  and in general we have
\begin{equation}\label{transv2}
  V_2=\int\sqrt{q}\td^2x=\int e^{2\Phi}\sqrt{\gamma^{(\tilde{k})}}\td^2y\neq\int\sqrt{\gamma^{(\tilde{k})}}\td^2y\,.
\end{equation}
We now define
\begin{equation}\label{conformeq2b}
  k:=\tilde{k}V_2^{-1}\int\sqrt{\gamma^{(\tilde{k})}}\td^2y\,.
\end{equation}
The value of $k$ is constant and may be different from $\tilde{k}$. However, it is clear that $k\geq0$ and $k=0$ iff $\tilde{k}=0$. Eq.~\eqref{aurcompteq8b} then leads to
\begin{equation}\label{aurcompteq8c}
\begin{split}
&\frac{\td}{\td z}(z^3F)=\frac{3e^{2\beta}}{z^4\el^2}+\frac{ke^{2\beta}}{z^2}\\
&+\frac{1}{z^2V_2}\int e^{2\beta}[(\hat{\pD}\beta)^2-\hat{\pD}^2{\Phi}-\gamma^{(\tilde{k})}_{ij}\tilde{A}^i\tilde{A}^j-Q^2]\sqrt{\gamma^{(\tilde{k})}}\td^2y\,.
\end{split}
\end{equation}
As we require that the spacetime is asymptotically planar/Schwartzchild AdS black hole, the function $\beta, A^i$ and $Q$ should decay to zero fast enough when $z\rightarrow0$. Thus, near the AdS boundary, Eq.~\eqref{aurcompteq8c} reduces into
\begin{equation}\label{aurcompteq8d}
\frac{\td}{\td z}(z^3F)=\frac{3}{z^4\el^2}+\frac{k}{z^2},~~z\rightarrow0\,,
\end{equation}
which gives us solution
\begin{equation}\label{z0solueq1}
  F(z)=-\left[kz^2+\frac1{\el^2}-f_0z^3\right]z^{-6},~~z\rightarrow0\,.
\end{equation}
Here the integration constant $f_0$ give us the mass according to Eq.~\eqref{defMass}. For finite $z$, Eq.~\eqref{aurcompteq8c} leads to following inequality
\begin{equation}\label{aurcompteq8e}
\begin{split}
&\frac{\td}{\td z}(z^3F)\leq\frac{3}{z^4\el^2}+\frac{k}{z^2}\\
&+\frac{1}{z^2V_2}\int e^{2\beta}[(\hat{\pD}\beta)^2-\hat{\pD}^2{\Phi}-\gamma^{(\tilde{k})}_{ij}\tilde{A}^i\tilde{A}^j-Q^2]\sqrt{\gamma^{(\tilde{k})}}\td^2y\,.
\end{split}
\end{equation}
Here we have used the facts $\beta\leq0$ and $k>0$. Integrating it with respect to $z$ and noting the value of $F(z)$ as $z\rightarrow0$, we obtain
\begin{equation}\label{solutildeR2}
  F(z)\leq-[kz^{2}+1/(\el^2)-f_0z^3]z^{-6}+\frac{B_z}{z^3V_2}\,.
\end{equation}
Here we define
\begin{equation}\label{eqforBz3}
\begin{split}
  B_{\tilde{z}}&=\int_0^{\tilde{z}}\frac{\td z}{z^2}\int e^{2\beta}[(\hat{\pD}\beta)^2-\hat{\pD}^2{\Phi}\\
  &-\gamma^{(\tilde{k})}_{ij}\tilde{A}^i\tilde{A}^j-Q^2]\sqrt{\gamma^{(\tilde{k})}}\td^2y\,.
  \end{split}
\end{equation}
Here $\hat{\pD}_i$ is the covariant derivative operator of $\gamma_{ij}$, $\tilde{A}^i$ is $A^i$ in the conformal frame. If all functions involved in the functional $B_{\tilde{z}}$ are free, $B_{\tilde{z}}$ has no upper bound or lower bound. However, these functions are not free. Due to Einstein's equation, spacetime geometry is determined by the distribution of matters. Because of Bianch identity and gauge $\partial_zq=0$, there are 5 bulk degrees of freedom. In fact we can use arbitrary 5 independent bulk variables. Here we choose $\{\Phi, W=\sqrt{T_{zz}},\tilde{A}^i,Q\}$ as 5 independent variables, i.e. $B_z=B_z[\Phi, W, \tilde{A}^i, Q]$. It needs to note that function $\beta$ depends on $\{\Phi, W, \tilde{A}^i, Q\}$ and is not a free variable. In following we will use variational method to show $B_z\leq0$.

We first consider an enlightening example, saying a smooth function $f(x)$ with $x\in(\infty,\infty)$. We can use two steps to prove $f(x)\leq a$: (i) $f(x)\leq a$ as $|x|\rightarrow\infty$ and (ii) for arbitrary saddle point $x_i$, i.e. the point of $f'(x_i)=0$, we have $f(x_i)\leq a$. This method can be generalized into functional case. If $\{||\Phi||, ||W||,||\tilde{A}^i||,||Q||\}\rightarrow\infty$ (here $||\cdot||$ is a $L^p$ norm), the system will break the spherical or planar symmetry strongly, which implies $|\partial_zq_{ij}|,T_{zz}, |(\hat{\pD}\beta)^2|$ and $|\hat{\pD^2}\Phi|\sim\mathcal{O}(\text{polynomial of }N)$ with a parameter $N\gg1$. The Eq.~\eqref{uucompteq} implies $e^{2\beta}\sim\exp[-\mathcal{O}($polynomial of $N)]$, which implies that
\begin{equation}\label{largeN1}
  \int e^{2\beta}[(\hat{\pD}\beta)^2-\hat{\pD}^2{\Phi}]\sqrt{\gamma^{(\tilde{k})}}\td^2y\rightarrow0,~~\text{as}~N\rightarrow\infty\,.
\end{equation}
Then we see $B_z\leq0$ as $N\rightarrow\infty$ by using the definition~\eqref{eqforBz3}.  This finishes the first step. In the second step, we use variational method to find all saddle points. The variation with respect to $\Phi$ shows $\hat{\pD}^2e^{2\beta}=0$, which implies
\begin{equation}\label{varbeta1}
  0=\int e^{2\beta}\hat{\pD}^2e^{2\beta}\sqrt{\gamma}\td^2y=\int(\hat{\pD}e^{2\beta})^2\sqrt{\gamma^{(\tilde{k})}}\td^2y
\end{equation}
and so $\beta$ is independent of $y^i$. Take this into integration~\eqref{eqforBz3} and we find on-shell value $B_z$
\begin{equation}\label{eqforBz4}
\begin{split}
  &B_{\tilde{z}}|_{\text{on-shell}}\\
  =&-\int_0^{\tilde{z}}\frac{\td z}{z^2}e^{2\beta}\int(\hat{\pD}^2{\Phi}+\gamma^{(\tilde{k})}_{ij}\tilde{A}^i\tilde{A}^j+Q^2)\sqrt{\gamma^{(\tilde{k})}}\td^2y\\
  =&-\int_0^{\tilde{z}}\frac{\td z}{z^2}e^{2\beta}\int(\gamma^{(\tilde{k})}_{ij}\tilde{A}^i\tilde{A}^j+Q^2)\sqrt{\gamma^{(\tilde{k})}}\td^2y\leq0\,.
  \end{split}
\end{equation}
Thus, we find $B_z\leq0$.  Then we obtain Eqs.~\eqref{solutildeR} and \eqref{upperforR} again for the case $d=3$.



There is also a rigidity theorem for 3+1 dimensional asymptotically plana/spherically AdS black hole: the inequality~\eqref{conjbd1} is saturated if and only if its geometry outside maximal cross-section is planar Schwarzschild-AdS. The proof is as follow.

To reach the maximum, we need to saturate Eqs.~\eqref{pertub2}, \eqref{aurcompteq8e} and \eqref{eqforBz4}. To saturate Cauchy-Schwartz inequality~\eqref{pertub2}, we need
\begin{equation}\label{rigeq1}
  \partial_i(e^{2\beta}\sqrt{q})=\partial_i(\tf\sqrt{q})=0\,,
\end{equation}
i.e. $e^{2\beta}\sqrt{q}$ and $\tf\sqrt{q}$ are independent of $x^i$. To saturate Eq.~\eqref{aurcompteq8e}, we have to set
\begin{equation}\label{rigeq2}
  \beta=0\Rightarrow\partial_i\beta=\partial_z\beta=0\,.
\end{equation}
Then Eq.~\eqref{rigeq1} implies $\partial_iq=0$ and so
\begin{equation}\label{rigeq3}
  \tf=\tf(z)\,.
\end{equation}
Combining Eq.~\eqref{rigeq2} and \eqref{uucompteq}, noting dominate energy condition requires $T_{zz}\geq0$, we find
\begin{equation}\label{rigeq4}
  \partial_zq_{ij}=0\,.
\end{equation}
This shows that $\R$, which is the scalar curvature of $q_{ij}$ is independent of $z$.
To saturate Eq.~\eqref{eqforBz4} we have to set $A^i=Q=0$, which implies
\begin{equation}\label{rigeq4}
  U^i=0\,.
\end{equation}
Then take Eqs.~\eqref{rigeq2}-\eqref{rigeq4} into Eq.~\eqref{urcompteq4} and we find
\begin{equation}\label{urcompteq4b}
  -2z^2\partial_z[z^{-3}{\tf(z)}]=\R+\frac{6}{z^2\el^2}\,,
\end{equation}
so we see $\R$ is only function of $z$ but independent of $x^i$. However, we have know $\R$ is independent of $z$, so $\R$ is constant. Then we find $\R=2k\geq0$ and Eq.~\eqref{urcompteq4b} shows $\tf=kz^2+1/\el^2-f_0z^3$.  The metric outside the horizon
\begin{equation}\label{metricBS1s1}
  \td s^2=\frac1{z^2}[-\mathfrak{f}(z)\td u^2+2\td u\td z+q_{ij}(x)]\td x^i\td x^j\,.
\end{equation}
As the 2-d metric $q_{ij}(x)$ has constant curvature $\R=2k$, we can always find suitable coordinates transformation $x^i\rightarrow\tilde{x}^i$ so that $q_{ij}$ becomes standard metric of sphere ($k>0$) or plane ($k=0$). Thus, we show that its geometry outside maximal cross-section is Schwarzschild-AdS with planar or spherical symmetry. On the other hand, we have known that the planar Schwarzschild-AdS black hole can saturate the bound but the spherical Schwarzschild-AdS black hole cannot. Thus, we prove our rigidity theorem. It needs to note that, there is no restriction on the geometry behind the maximal cross-section.

\section{Summary and discussion}
To conclude, this paper proposed and discussed a new universal inequality for the inner geometry of black holes. This makes a first step towards the holographic proof on the conjecture that vacuum black holes may be fastest ``computers'' in nature~\cite{Brown:2015bva,Brown:2015lvg}. Except for seeking the proof about Eq.~\eqref{conjbd1} in more general cases, many other aspects are worthy of exploring in the future.

In the proofs of this paper, it is crucial that scalar curvature of $\Gamma_z$ is nonnegative when $z\rightarrow\infty$. This is why it requires that the spacetime is asymptotically planar/spherical Schwarzschild-AdS. The bound~\eqref{conjbd1} can be violated by asymptotically hyperbolic black holes. In fact hyperbolically AdS black hole can have negative energy so Eq.~\eqref{conjbd1} is not true. It is interesting to study if there is other suitable upper bound for hyperbolic case. In our above discussions and proofs, we only consider Einstein theory. It would be also interesting to consider other gravity theories.

Assume that there is a next-outermost horizon $\mathcal{H}_2$ behind the outermost horizon $\mathcal{H}_1$. In the limit $\mathcal{H}_2\rightarrow\mathcal{H}_1$, i.e., the temperature $T_H\rightarrow0$, $\max\Sigma\rightarrow0$ but the total energy can be arbitrarily large. This suggests that, in low temperature limit, there may be an tighter upper bound controlled by temperature. For example, in a BTZ black hole~\eqref{metricBTZ}, the function $f$ can be rewritten in terms of
\begin{equation}\label{fz1z2}
  f=\frac{(z^2-z_1^2)(z^2-z_2^2)}{z_1^2z_2^2\el^2}\,.
\end{equation}
Here $z_1\leq z_2$ are inverse radii of horizons. Then we see $f_0=(z_1^2+z_2^2)/(z_1^2z_2^2\el^2)$ and $J=2/(z_1z_2\el)$. Then we see Eq.~\eqref{btzsig2} becomes
\begin{equation}\label{btzsig2b}
\begin{split}
  \max\Sigma=V_1\sqrt{\el^2f_0^2/4-J^2/4}=V_1\frac{z_2^2-z_1^2}{2z_2^2z_1^2\el}\,.
  \end{split}
\end{equation}
On the other hand, the temperature $T_H$ and entropy $S$ can be expressed as
\begin{equation}\label{btzts1}
  T_H=\frac{z_2^2-z_1^2}{2\pi\el^2 z_2^2z_1},~~S=V_1/z_1\,.
\end{equation}
Then one can verify $\max\Sigma= \pi T_H S\el$ and so we obtain a new bound for BTZ black hole
\begin{equation}\label{newbdbtz}
  \Sigma\leq\pi T_H S\el\,.
\end{equation}
In the low temperature limit, this bound is much tighter than inequality~\eqref{conjbd1}. It is interesting to check if a similar bound is also true in general cases.


The bound can be generalized into asymptotic flat spacetimes. It only needs to set $k=1$ and $\el\rightarrow\infty$. This leads to a different bound $\Sigma\leq c_dE^{(d-1)/(d-2)}$ with a dimension-dependent number $c_d$. One find this result from Eq.~\eqref{solutildeR}. By setting $k=1$ and $\el\rightarrow\infty$, we see that
\begin{equation}\label{solutildeRflat2}
  F(z)\leq-z^{-2d}(z^2-f_0z^d)\,.
\end{equation}
The maximal value of right-hand in Eq.~\eqref{solutildeRflat2} is proportional to $f_0^{2(d-1)(d-2)}$, so Eq.~\eqref{ineqFs1} implies
\begin{equation}\label{upperforR0}
  \max\Sigma\leq c_dE^{(d-1)/(d-2)}\,.
\end{equation}
Particularly, in four-dimensional spacetime, we have $\Sigma\leq 3\sqrt{3}\pi^2E^2$.  The first evidence of this inequality was reported by Ref.~\cite{Christodoulou:2014yia} based on the comparsion between Schwarzschild black hole and Reissner-Nordstr\"{o}m black hole. This inequality gives us a way to estimate the possible largest volume inside black hole. For a black hole formed by collapse, at the time $\tau$ much later than the disappearance of the collapsing matter, $(\partial/\partial t)^I$ is Killing vector approximately, so we find that spatial volume attached at the horizon is bounded by $V(\tau)\lesssim \tau\times\max\Sigma= 3\sqrt{3}\pi^2E^2 \tau$.  For a large evaporating black hole of energy $E$, the black hole will be quasi-stationary in a time order $\mathcal{O}(E^3)$ and so the maximal volume of space inside horizon is order $\mathcal{O}(E^5)$.

\begin{acknowledgments}
The author would like to thank Qi Wang for her encourage and support.
\end{acknowledgments}

\appendix

\section{Size of cross-section in coordinates gauge \eqref{metricflat1}}\label{matrix1}
Let us first explain how to obtain Eq.~\eqref{defLRL}. For the nontrivial cross-section $\mathcal{S}_{d-1}$, we can always embed it into the a $d$-dimensional spacelike manifold $\mathcal{M}_d$ of which unite normal covector $\tilde{n}_I$ satisfies $\tilde{n}_I|_{\mathcal{S}_{d-1}}=n_I$. In following, we will use $x^I=\{t,x^i,z\}$ to stand for $d+1$ dimensional bulk coordinates, $x^\mu=\{t,x^i\}$ to stand for the coordinates of $d$-dimensional space-like submanifold $\mathcal{M}_d$ and $x^i$ to stand for $d-1$ dimensional coordinates of cross-section $\mathcal{S}_{d-1}$.

We parameteriz the surface $\mathcal{M}_{d}$ to be $z=z(t,x^i)$. The normal covector reads
\begin{equation}\label{Mcovectorn1}
  \tilde{n}_I\propto(\td z)_I-\dot{z}(\td t)_I-(\partial_iz)(\td x^i)_I\,.
\end{equation}
The condition $\tilde{n}_I|_{\mathcal{S}_{d-1}}=n_I$ and $\xi^In_I=(\partial/\partial t)^In_I=0$ implies
\begin{equation}\label{condSdz1}
  \dot{z}|_{\mathcal{S}_{d-1}}=0\,.
\end{equation}
Here dot means the partial derivative with respect to $t$. The induced metric on this surface then reads,
\begin{equation}\label{metricinduce}
\begin{split}
  &\td s^2_d=\frac1{z^2}[-(f+\chi\dot{z}^2)\td t^2+2\chi\dot{z}\partial_iz\td t\td x^i\\
  &2v_i\td t\td x^i+(h_{ij}+\chi\partial_iz\partial_jz)\td x^i\td x^j]
  \end{split}\,.
\end{equation}
The choice for such $d$-dimensional manifold is not unique. We can choose $z=z(t,x^i)=z_{\mathcal{S}}(x^i)$ and so the metric of $\mathcal{M}_d$ becomes
\begin{equation}\label{metricinduce2}
\begin{split}
\td s^2_d=&\frac1{z^2}[-f\td t^2+2v_i\td t\td x^i+\tilde{h}_{ij}\td x^i\td x^j]\,.
\end{split}
\end{equation}
Here $\tilde{h}_{ij}=h_{ij}+\chi\partial_iz\partial_jz$.
In this choice the Killing vector $\xi^I$ lays in $\mathcal{M}_d$ and is still a Killing vector of $\mathcal{M}_d$. In the sub-manifold $\mathcal{M}_d$, we denote this Killing vector to be $\xi^\mu$. The cross-section $\mathcal{S}_{d-1}$ then is given by a hypersurface $t=t_{\mathcal{S}}(x^i)$ in $\mathcal{M}_d$. Assume that $\td S_\mu$ is the directed surface element of cross-section $\mathcal{S}_{d-1}$ embedded in $\mathcal{M}_d$. The size of cross-section then reads
\begin{figure}
  \centering
  \includegraphics[width=.5\textwidth]{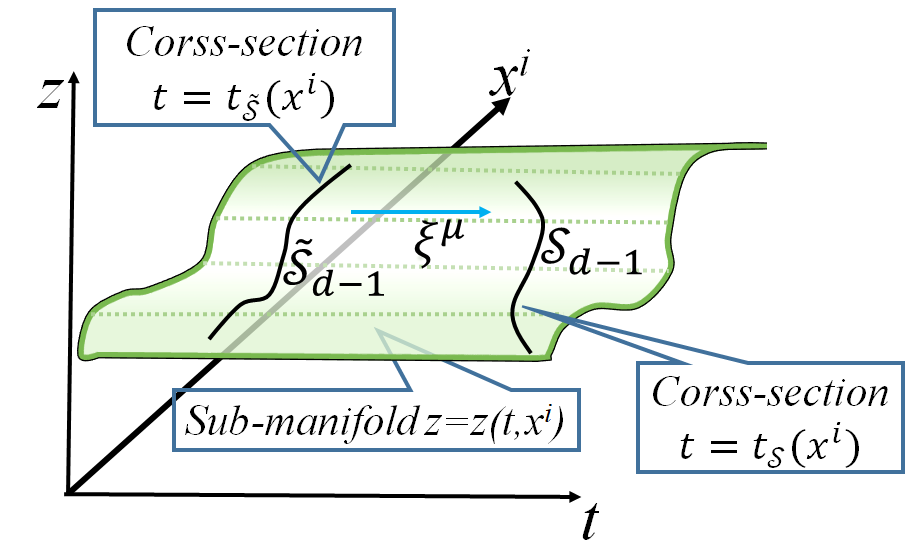}
  \caption{The schematic diagram about the submanifold $\mathcal{M}_d$, cross-section $\mathcal{S}_{d-1}$ and cross-section $\tilde{\mathcal{S}}_{d-1}$. $\xi^\mu=(\partial/\partial t)^\mu$ is the Killing vector which lays in $\mathcal{M}_d$.
  }
   \label{CVinSAdS2}
\end{figure}
\begin{equation}\label{newsigma1}
  \Sigma[\mathcal{S}_{d-1},\xi^I]=\int_{\mathcal{S}_{d-1}}\xi^\mu\td S_\mu\,.
\end{equation}

We consider a new cross-section $\tilde{S}_{d-1}$, which is given by $t=t_{\tilde{\mathcal{S}}}(x^i)$ in $\mathcal{M}_d$. See the schematic diagram Fig.~\ref{CVinSAdS2}. Then we have
\begin{equation}\label{newsigma1}
  \Sigma[\tilde{\mathcal{S}}_{d-1},\xi^I]=\int_{\tilde{\mathcal{S}}_{d-1}}\xi^\mu\td S_\mu\,.
\end{equation}
Using the Causs formula, we have
\begin{equation}\label{newsigma1}
  \Sigma[\mathcal{S}_{d-1},\xi^I]-\Sigma[\tilde{\mathcal{S}}_{d-1},\xi^I]=\int_{\Xi_d}D_\mu\xi^\mu\td V\,.
\end{equation}
Here $\Xi_d$ is the $d$-dimensional region surrounded by $\tilde{\mathcal{S}}_{d-1}$ and $\mathcal{S}_{d-1}$, $D_\mu$ is the covariant derivative operator of sub-manifold $\mathcal{M}_d$. As $\xi^\mu$ is a Killing vector of $\mathcal{M}_d$, $\xi^\mu$ is divergent free $D_\mu\xi^\mu=0$. Then we find
$$\Sigma[\mathcal{S}_{d-1},\xi^I]=\Sigma[\tilde{\mathcal{S}}_{d-1},\xi^I]\,.$$
Thus we proved that the value of $\Sigma[{\mathcal{S}}_{d-1},\xi^I]$ is independent of the choice of $t_{{\mathcal{S}}}(x^i)$. Particularly, we can choose $t_{\tilde{\mathcal{S}}}(x^i)=0$ and so
\begin{equation}\label{normalS1}
  \td S_\mu=\sqrt{-f-|v|^2}(\td t)_\mu\sqrt{\tilde{h}}\td^{d-1}x\,.
\end{equation}
Here $|v|^2:=\tilde{h}^{ij}v_iv_j$. Thus,
\begin{equation}\label{newsigma2}
\begin{split}
  &\Sigma[\mathcal{S}_{d-1},\xi^I]=\Sigma[\tilde{\mathcal{S}}_{d-1},\xi^I]=\int_{\tilde{\mathcal{S}}_{d-1}}\xi^\mu\td S_\mu\\
  =&\int_{z=z_{\mathcal{S}}(x^i)}z^{-d}\sqrt{-f-|v|^2}\sqrt{\tilde{h}}\td^{d-1}x\,.
  \end{split}
\end{equation}
Thus, we obtain Eq.~\eqref{defLRL}.

\section{Size of cross-section in Bondi-Sachs gauge}\label{bfg1}
If we use the Bondi-Sachs coordinates gauge, the spacelike hypersurface $\mathcal{M}_d$ inside black hole then is parameterized by $z=z(u,x^i)$ and the induced metric reads
\begin{equation}\label{metricBSind3a}
\begin{split}
  \td s^2=&\frac1{z^2}[-({\tf}-2\dot{z})e^{2\beta}\td u^2+2e^{2\beta}\partial_iz\td u\td x^i\\
  &+q_{ij}(\td x^i-U^i\td u)(\td x^j-U^j\td u)]]
  \end{split}
\end{equation}
Here dot means the partial derivative with respect to $u$. The normal covector of $\mathcal{M}_d$ reads
\begin{equation}\label{Mcovectorn1}
  \tilde{n}_I\propto(\td z)_I-\dot{z}(\td u)_I-(\partial_iz)(\td x^i)_I\,.
\end{equation}
The condition $\tilde{n}_I|_{\mathcal{S}_{d-1}}=n_I$ and $\xi^In_I=(\partial/\partial u)^In_I=0$ implies
\begin{equation}\label{condSdz1}
  \dot{z}|_{\mathcal{S}_{d-1}}=0\,.
\end{equation}
The choice for such $d$-dimensional manifold is also not unique. We can choose $z=z(u,x^i)=z_{\mathcal{S}}(x^i)$ and so the metric of $\mathcal{M}_d$ becomes
\begin{equation}\label{metricBSind3b}
\begin{split}
  \td s^2=&\frac1{z^2}[(|U|^2-{\tf}e^{2\beta})\td u^2+2(e^{2\beta}\partial_iz-U_i)\td u\td x^i\\
  &+q_{ij}\td x^i\td x^j]\,.
  \end{split}
\end{equation}
Here $|U|^2=U^iU_i$ and $U_i=q_{ij}U^j$. In this choose, the $(\partial/\partial u)^I$ is still tangent to $\mathcal{M}_d$ and so is the Killing vector of $\mathcal{M}_d$.

In the sub-manifold $\mathcal{M}_d$, the cross-section $\mathcal{S}_{d-1}$ then is given by a hypersurface $u=u_{\mathcal{S}}(x^i)$ in $\mathcal{M}_d$. Similar to the former case, we can find that the size of cross-section is independent of the choice of $u_{\mathcal{S}}(x^i)$ and we obtain
\begin{equation}\label{newsigma2}
\begin{split}
  &\Sigma[\mathcal{S}_{d-1},\xi^I]\\
  =&\int_{z=z_{\mathcal{S}}(x^i)}\frac{e^{\beta}}{z^d}\sqrt{-\tf-e^{2\beta}|\partial z|^2+2U^i\partial_iz}\td V_{d-1}\,.
  \end{split}
\end{equation}
Here $\td V_{d-1}:=\sqrt{q}\td^{d-1}x$ and $|\partial z|^2=q^{ij}\partial_iz\partial_jz$.

\bibliographystyle{JHEP}
\bibliography{CV-bound}

\end{document}